\documentclass[12pt]{article}

\input{myCommands.sty}

\newcommand{\blind}{0}

\begin{document}

\def\spacingset#1{\renewcommand{\baselinestretch}%
{#1}\small\normalsize} \spacingset{1}



\if0\blind
{
  \title{\bf Autoregressive Density Modeling with the Gaussian Process Mixture Transition Distribution}
  \author{Matthew Heiner and Athanasios Kottas \thanks{
    M. Heiner is Assistant Professor, Department of Statistics,
    Brigham Young University, and A. Kottas is Professor, Department of Statistics, University of California, Santa Cruz. This work is part of the Ph.D. dissertation of the first author, completed at the University of California, Santa Cruz.}
    \hspace{.2cm}}
  \maketitle
} \fi

\if1\blind
{
  \bigskip
  \bigskip
  \bigskip
  \begin{center}
    {\LARGE\bf Title}
\end{center}
  \medskip
} \fi

\bigskip
\begin{abstract} 

  We develop a mixture model for transition density approximation, together with soft model selection, in the presence of noisy and heterogeneous nonlinear dynamics. Our model builds on the Gaussian mixture transition distribution (MTD) model for continuous state spaces, extending component means with nonlinear functions that are modeled using Gaussian process (GP) priors. The resulting model flexibly captures nonlinear and heterogeneous lag dependence when several mixture components are active, identifies low-order nonlinear dependence while inferring relevant lags when few components are active, and averages over multiple and competing single-lag models to quantify/propagate uncertainty. Sparsity-inducing priors on the mixture weights aid in selecting a subset of active lags. The hierarchical model specification follows conventions for both GP regression and MTD models, admitting a convenient Gibbs sampling scheme for posterior inference. We demonstrate properties of the proposed model with two simulated and two real time series, emphasizing approximation of lag-dependent transition densities and model selection. In most cases, the model decisively recovers important features. The proposed model provides a simple, yet flexible framework that preserves useful and distinguishing characteristics of the MTD model class.

\end{abstract}

\noindent%
{\it Keywords}: Time Series, Lag Selection, Nonlinear Autoregression, Dynamical System, Time-delay Embedding  
\vfill

\newpage
\spacingset{1.5} 

\section{Introduction}
\label{sec:intro}

In this article, we propose and explore a mixture model for transition density approximation, as well as for soft model selection via shrinkage, in the presence of noisy and heterogeneous nonlinear dynamics. Common uses of mixture modeling techniques include accounting for unexplained heterogeneity, robustness to outliers, complex density estimation through convolution, model-based clustering and deconvolution, and model averaging, among others. These objectives have found increased use in time series analysis for capturing local or time-dependent dynamics \citep{carvalho2005,wood2011mixAR}, modeling complex transition densities \citep{deyoreo2017}, and combining ensembles of forecasts \citep{raftery2005weather}, etc.; see \citet{fruhwirth2006} for a comprehensive introduction. One pioneering method in this area is the class of mixture transition distribution (MTD) models of \citet{raftery1985}. Originally proposed as a parsimonious approximation to a higher-order Markov chain, the MTD model leverages a mixture of univariate conditional distributions, each using a distinct input. Integrating these simple parts into a global model has provided a useful framework with several applications \citep{berchtold2002, hassan2006modeling, hassan2013fitting,escarela2006flexible, luo2009parameter}. Our contribution builds on the continuous-state MTD model of \citet{le1996gmtd}.  %

\cite{martin1987comment} were the first to observe that the MTD framework extends beyond discrete state spaces.
The general MTD formulation for the conditional distribution $F$ on time series $\{ y_t \}_{t=1}^T \in \mathbb{R}^T$ is given by
\begin{align}
\label{eq:mtd_general}
    F_t(y_t \mid y_{t-1}, \ldots, y_1) = \sum_{\ell = 1}^L \lambda_\ell \, G_\ell( y_t \mid y_{t-\ell} ) \, ,
\end{align}
where each mixture component contains a univariate transition law $G_\ell$ associated with a specific lag, up to a fixed horizon $L$, and mixing weights $\lambda_\ell \ge 0$ with $\sum_{\ell=1}^L \lambda_\ell = 1$. Although the general model in (\ref{eq:mtd_general}) is most fundamentally a mixture, its structure resembles widely used linear autoregressive models \citep{berchtold2002}. 
In this sense, MTD models fit into the mainstream of methods tailored to continuous state spaces more than they do in their originally proposed domain of discrete Markov chains.

The most popular, and perhaps simplest model belonging to the continuous-state MTD family is the Gaussian MTD (GMTD) proposed by \citet{le1996gmtd}, wherein $G_\ell$ corresponds to a Gaussian distribution with linear mean $\beta_\ell \, y_{t-\ell}$ and variance $\sigma_\ell^2$. Their model further includes a mixture component containing a full linear autoregressive (AR) model of order $L$. With this simple form, the GMTD offers flexibility and better captures characteristics unavailable to standard AR models. Further modifications include a zero-mean component with large variance to accommodate outliers, and a random-walk specification to accommodate flat stretches. 

Despite this flexibility to model what are often termed as ``nonlinear'' time series, the GMTD model is restricted to have a linear and additive transition mean. If we denote the coefficients for the full AR component as $\beta_{01}, \ldots, \beta_{0 L}$, then the conditional transition mean for the GMTD is
$\E(y_t \mid y_{t-1}, \ldots, y_1) = \sum_{\ell=1}^L ( \lambda_0 \, \beta_{0 \ell} + \lambda_\ell \, \beta_\ell ) \, y_{t-\ell}$. While this linear structure is important for deriving stationarity conditions, we forego this restriction in favor of estimating nonlinear dependence. Thus we will consider each $G_\ell(y_t \mid y_{t-\ell})$ to have a separate location $ \mu_\ell + f_{\ell}(y_{t - \ell}) $ consisting of a level and continuous nonlinear function $f_\ell(y_{t-\ell})$ mapping the relevant lag to $\mathbb{R}$. If the location of $G_\ell$ also represents the conditional mean, then we have $\E(y_t \mid y_{t-1}, \ldots, y_1) = \mu + \sum_{\ell=1}^L \lambda_\ell \, f_\ell(y_{t-\ell})$ where $\mu = \sum_{\ell=1}^L \lambda_\ell \, \mu_\ell$. This resembles the conditional mean obtained from the popular generalized additive model family \citep{hastie1990GAMbook}, which has been applied to autoregressive models \citep{chen1993GAMAR, wong1996bayesGAMAR}. \citet{huang2004GAMARorder} further consider order selection using the Bayesian information criterion (BIC) in this context. As \citet{le1996gmtd} note, however, the MTD formulation is distinct from generalized additive models in that the errors arise from a mixture. That is, rather than averaging surfaces into a single composite with homogeneous error, the MTD model uses the functions $f_\ell$ to define the error mixture, which can vary widely across the input space.


Instead of focusing on approximation of a multi-dimensional surface, our primary objectives include: 1) flexibly modeling nonlinear and heterogeneous lag dependence when several mixture components are active, 2) identifying low-order nonlinear dependence while inferring relevant lags when few components are active, and 3) averaging over multiple and competing single-lag models to more appropriately quantify/propagate uncertainty. 
To provide nonlinearity, we model the unknown functions $\{ f_\ell \}$ with Gaussian process (GP) priors, which have been applied extensively for time series (see \citealp{gregorcic2009GPnonlineardynamics, kocijan2003predictivecontrolGP, gutjahr2012sparseGPtimeseries} for examples in the nonlinear autoregressive context) and nonlinear regression generally (see \citealp{rasmussen2006gp}, and references therein), including mixture modeling \citep{shi2003mixGP} and generalized additive formulations \citep{duvenaud2011additiveGP}. By heterogeneity, we mean that the transition density flexibly adapts across the multidimensional lag space, capturing heteroscedastic and multimodal behaviors. This is accomplished through the mixture, which simultaneously provides model averaging and/or selection through continuous shrinkage of the mixture weights.

The rest of the article is organized as follows. In Section \ref{sec:model}, we develop the proposed model, which we call the Gaussian-process mixture transition distribution (GPMTD), and discuss Bayesian estimation and forecasting. In Section \ref{sec:illustrations}, we demonstrate the model with simulated and real time series, highlighting both the model selection and model averaging strengths of the proposed methodology. We conclude with discussion in Section \ref{sec:discussion}. Additional technical details are provided in the appendices.

\section{The modeling approach}
\label{sec:model}

We propose a model with basic form
\begin{align}
    \label{eq:gpmtd1}
    F_t(y_t \mid y_{t-1}, \ldots, y_1) &= \lambda_0 \, \Ndist\left( y_t \mid \mu_0, \sigma_0^2 \right) + \sum_{\ell = 1}^L \lambda_\ell \, \Ndist \left( y_t \mid \mu_\ell + f_\ell(y_{t-\ell}), \, \sigma_\ell^2 \right) \nonumber \\ 
    f_\ell &\simindep \GP, \quad \bm{\lambda} \sim p(\bm{\lambda}) \, , 
\end{align}
where $\Ndist(\cdot \mid \mu, \, \sigma^2)$ corresponds to a univariate Gaussian distribution with mean $\mu$ and variance $\sigma^2$, and $\bm{\lambda} = (\lambda_0, \lambda_1, \ldots, \lambda_L)$. To aid in the selection objective, we employ a specialized prior $p(\bm{\lambda})$, described in Section \ref{sec:SBM}, that admits soft sparsity through shrinkage, jumps to effectively omit inactive lags, and stochastic ordering of active lags to reflect the common belief that recent lags generally carry greater influence. To emphasize the MTD structure and our chosen objectives, we drop the full AR mixture component present in the GMTD. We retain the zero-indexed component (intercept) in order to accommodate what \citet{le1996gmtd} term replacement-type outliers, to add flexibility to the mixture, and to contribute to a stationary distribution in the absence of serial dependence.

The hierarchical specification for the GPMTD model follows standard conventions for both Gaussian process regression and MTD models. To facilitate computation, we break the mixture with latent component membership indicators for each time point, $z_t \in \{0,1,\ldots,L\}$. To distinguish $y_t$ from its lags and to emphasize that covariates could be incorporated into the framework, we denote the time-delay vector as $\bm{x}_t = (x_{t,1}, \ldots, x_{t,L}) \equiv (y_{t-1}, \ldots, y_{t-L})$. For the Gaussian process priors, we focus on common default choices of Mat\'ern covariance functions with Euclidean distance and a smoothness parameter $\nu$ fixed at 2.5 or $+\infty$, the former value ensuring a twice-differentiable regression function $f$ and the latter corresponding to the squared exponential covariance function \citep{rasmussen2006gp}.

Treating the first $L$ observations of the time series as fixed, and implicitly conditioning the top level on lags in $\bm{x}_t$, the full hierarchical representation for the model in (\ref{eq:gpmtd1}) is given by
\begin{align}
    \label{eq:hier}
    y_t \mid 
    z_t, \mu_0, \sigma_0^2, \{ (\mu, \sigma^2, f)_\ell \}_{\ell=1}^L &\simindep 
    \begin{cases}
        \Ndist \left(\mu_0, \sigma_0^2 \right) & \text{if} \ z_t = 0, \\
        \Ndist \left(\mu_\ell + f_\ell(x_{t,\ell} \right), \sigma_\ell^2) & \text{if} \ z_t = \ell \in \{1, \ldots, L \} \, ,
    \end{cases} \nonumber \\
    & \qquad  \text{for} \ t = L+1, \ldots, T, \nonumber \\
    \Pr(z_t = \ell \mid \bm{\lambda}) = \lambda_\ell \, , \ \text{for} \ \ell &= 0,1,\ldots,L, \ \text{independently for} \ t=L+1,\ldots,T, \nonumber \\
    \bm{\lambda} &\sim \SBM(\eta_\lambda, \pi_{\lambda,1}, \pi_{\lambda,3}, \bm{\gamma}_\lambda, \bm{\delta}_\lambda), \\
    \mu_\ell \simindep \Ndist\left(m_0^{(\ell)}, v_0^{(\ell)}\right), \ \sigma_\ell^2 &\simindep \IGdist \left( \nu_\sigma^{(\ell)}/2, \nu_\sigma^{(\ell)}s_0^{(\ell)}/2 \right) , \quad \text{for} \ \ell = 0,1,\ldots,L, \nonumber \\
    f_\ell \mid \kappa_\ell, \sigma_\ell^2, \nu, \psi_\ell &\simindep \GP \left( \bm{0}, \kappa_\ell \, \sigma_\ell^2 \, \rho(x, x'; \nu, \psi_\ell) \right) ,  \quad \text{for} \ \ell = 1,\ldots,L, \nonumber \\
    \kappa_\ell \mid \nu_\kappa, \kappa_0 &\simindep \IGdist \left( \nu_\kappa/2, \nu_\kappa \, \kappa_0/2 \right) ,  \quad \text{for} \ \ell = 1,\ldots,L, \nonumber \\ 
    \psi_\ell \mid \nu_\psi, \psi_0 &\simindep \IGdist \left( \nu_\psi/2, \nu_\psi \, \psi_0/2 \right) ,  \quad \text{for} \ \ell = 1,\ldots,L, \nonumber \\
    p(\nu_\kappa) &\propto 1_{( \nu_\kappa \in \mathcal{V}_\kappa )}, \ \kappa_0 \sim \Gammadist(a_\kappa, b_\kappa), \nonumber \\
    p(\nu_\psi) &\propto 1_{( \nu_\psi \in \mathcal{V}_\psi )}, \ \psi_0 \sim \Gammadist(a_\psi, b_\psi), \nonumber
\end{align}
where the SBM is the stick-breaking mixture prior described in Section \ref{sec:SBM}; $\IGdist(a,b)$ denotes the inverse-gamma distribution with shape $a$ and scale $b$; the Gaussian process is characterized by the zero mean function denoted with $\bm{0}$ and covariance function $\kappa_\ell \,  \sigma_\ell^2 \, \rho(\cdot, \cdot; \nu, \psi_\ell)$ utilizing correlation function $\rho$ in the Mat\'ern class with smoothness parameter $\nu$ and length scale parameter $\psi$; $\mathcal{V}_\kappa$ and $\mathcal{V}_\psi$ are finite, discrete sets of positive real numbers; and $\Gammadist(c,d)$ denotes a gamma distribution with mean $c/d$. Each inverse-gamma distribution is parameterized in terms of a scaled inverse Chi-squared distribution with degrees of freedom and prior harmonic mean, which aid both with interpretation and computation (potentially reduced posterior correlation among the parameters). We also parameterize the GP variance as the product $\kappa  \, \sigma^2$ to aid with interpretation of $\kappa$ as a signal-to-noise ratio (SNR), as well as computation, obtaining a tractable collapsed conditional distribution for each $\sigma^2$ parameter. 

Because $\bm{x}_t$ contains lags of the time series, it may be reasonable to assume some degree of homogeneity among $\{ f_\ell \}$ across lags. We consequently allow hierarchical borrowing-of-strength in the parameters governing the covariance functions across $\ell = 1, \ldots, L$. Even with $\nu$ fixed, $\kappa$ and $\psi$ are not fully identified \citep{zhang2004maternid}, which further justifies our use of informative and hierarchically connected priors.

\subsection{Spike-and-slab prior for mixture weights}  
\label{sec:SBM}

We employ a sparsity-inducing prior on the mixture weights to aid in selecting a subset of active lags. The stick-breaking mixture (SBM) prior of \citet{heiner2019spv} has assisted with lag selection and identifiability in discrete MTD models \citep{heiner2019MMTDarXiv}, and we apply it 
similarly in the GPMTD model. The prior builds the probability vector $\bm{\lambda}$ through an extension of the stick-breaking construction that defines the generalized Dirichlet distribution \citep{connor1969}. In particular,
\begin{align}
\label{eq:stickbreaking}
\lambda_0 = \theta_0, \ \lambda_j = \theta_j \prod_{i=0}^{j-1} (1 - \theta_i) \ \text{for}\ j=1,\ldots,L-1, \ \text{and} \ \lambda_L = \prod_{i=0}^{L-1} (1 - \theta_i) \, ,
\end{align}
with $\theta_j$ independently drawn from a mixture of three beta distributions, $\theta_j \simindep \\ 
\pi_{\lambda,1} \Betadist(1, \eta_\lambda) + \pi_{\lambda, 2} \Betadist(\gamma_{\lambda, j}, \delta_{\lambda, j}) + \pi_{\lambda, 3} \Betadist(\eta_\lambda, 1)$, where $\pi_{\lambda,1} + \pi_{\lambda,2} + \pi_{\lambda,3} = 1$, $\bm{\gamma}_\lambda = (\gamma_{\lambda,0}, \gamma_{\lambda,1}, \ldots, \gamma_{\lambda,L-1})$, and $\bm{\delta}_\lambda = (\delta_{\lambda,0}, \delta_{\lambda,1}, \ldots, \delta_{\lambda,L-1})$ appear in (\ref{eq:hier}). One can use this mixture structure to encourage sparsity by setting $\eta_\lambda \gg 1$, in which case the first component corresponds to small probabilities in $\bm{\lambda}$. The third component allows for the rest of the unbroken stick (i.e., $\prod_{i=0}^{j-1}(1-\theta_i) = 1 - \sum_{i=0}^{j-1} \lambda_i$) to be used for $\lambda_j$, while the second mixture component allows for flexibility in modeling $\lambda_j$. The first and third mixture components could be thought of as providing spikes. In the second (slab) component, the $\gamma_{\lambda,j}$ and $\delta_{\lambda,j}$ parameters can be fixed at the same values across $j$, or can be set to mimic the Dirichlet distribution with a three-parameter extension. 
If the hyperparameters of the SBM prior are fixed, as is typical with Dirichlet priors, incorporating the SBM into a hierarchical model involving multinomial counts (latent or observed) requires minimal effort due to conditional conjugacy (see Appendix \ref{sec:appendix_SBM}). 

\subsection{Prior specification}

The GPMTD model is somewhat robust to prior choice, so long as the parameters governing variances are on an appropriate scale. Here, we provide some guidance and default prior values as a starting point. Experience simulating time series from the model suggests that values of the SNR parameter $\kappa$ on the order of $10^2$ or $10^3$ are necessary (when length scale $\psi$ is on the order of $1$) for smooth nonlinear dynamics to visually manifest.

We typically set the prior for the level parameters $\{\mu_\ell : \ell = 0, \ldots, L \}$ to have mean $m_\ell^{(0)} = 0$ and variance $v_0^{(\ell)}$ either large (one or two orders of magnitude greater than the range of the data) or commensurate with the range of the data. Absent strong beliefs about observation noise, we set all $\nu_\sigma = 5.0$ to ensure two finite moments in the inverse-gamma priors, with prior estimate $s_0^{(0)}$ large (approximately one order of magnitude greater than the range of the data) to accommodate outliers, and $s_0^{(\ell)} = 1.0$.

We employ informative hierarchical priors for all $\kappa$ (SNR) and $\psi$ (length scale) parameters. To simplify posterior sampling while allowing some flexibility, we use \\$\mathcal{V}_\kappa = \mathcal{V}_\psi = \{5.0, 7.5, 10.0, 25.0, 50.0\}$ to define default discrete uniform priors on the degrees of freedom (concentration) parameters of the inverse-gamma distributions. We also use as default values $a_\kappa=10.0$ and $b_\kappa=0.1$ for the gamma prior on $\kappa_0$ (the harmonic mean for each $\kappa$), yielding a prior mean of 100.0 for $\kappa_0$; and $a_\psi=10.0$ and $b_\psi=1.0$, yielding a prior mean of 10.0 for $\psi_0$ (the prior harmonic mean for each length scale $\psi$). If one has strong prior beliefs regarding the strength of the dynamic signal relative to observation noise, we recommend first carefully considering an informative prior for each $\sigma_\ell^2$ for $\ell > 0$, and then setting an informative prior for the $\kappa$ parameters by possibly increasing the values in $\mathcal{V}_\kappa$ and concentrating the gamma prior for $\kappa_0$.

The parameters in the SBM prior for the mixing weights should be thoughtfully considered in the context of each analysis, especially in cases with sample sizes $T < 50$. For example, priors overly concentrated on $\lambda_0$ in conjunction with a small $\sigma_0^2$ can result in an unintended bimodal transition distribution. \citet{heiner2019spv} provides guidance for selecting an SBM prior with a level of sparsity reflecting prior beliefs about the number of active lags in the time series. We employ default values of $\eta_\lambda = 1{,}000$, $\pi_{1\lambda} = 0.5$, $\pi_{3\lambda} = 0.25$, $\bm{\gamma}_\lambda = \bm{1}$, and $\bm{\delta}_\lambda = \bm{1}$ where $\bm{1}$ is a vector of ones. This results in a marginal prior density with peaks near the extremes and near-uniformity between 0 and 1 for each $\lambda_\ell$. 

\subsection{MCMC posterior simulation}
\label{sec:MCMC}

The hierarchical model in (\ref{eq:hier}) admits a convenient Gibbs sampling scheme for posterior inference. We highlight details unique to this model and outline the algorithm, deferring remaining details to Appendices \ref{sec:appendix_mixcomp_updates} and \ref{sec:appendix_SBM}. As is standard with Gaussian process regression, we work with the finite-dimensional distributions of the independent prior processes for $\{f_\ell\}$, which are multivariate Gaussian with mean 0 everywhere and covariance between all input pairs $(x,x'), \ x, x' \in \mathbb{R}$, parameterized as in (\ref{eq:hier}). Let $\bm{f}_\ell$ denote a length $T-L$ vector for which the $i$th element is the realization $f_{i,\ell} \equiv f_\ell(x_{i,\ell})$. To encourage mixing of the MCMC chain, we marginalize the full posterior over all $\{(\mu, \sigma, \bm{f})_\ell\}$ before updating $\{(\kappa, \psi)_\ell\}$, the only parameters for which collapsed/full conditional distributions are not tractable. For each $\ell = 1, \ldots, L$, we jointly update the pair $(\kappa, \psi)_\ell$ with a random-walk Metropolis step using bivariate Gaussian proposals on the logarithmic scale. Given these updates, conditionally conjugate updates are available for individual parameters in $\{(\mu, \sigma, \bm{f})_\ell\}$. We note that each $f_\ell$ must be evaluated at every $x_{t,\ell}$ (denoted as $f_{t,\ell}$) to facilitate full conditional draws for $\{z_t\}$, given as
\begin{align}
    \label{eq:zetaUpdate}
    \Pr(z_t = \ell \mid \cdots) = \frac{ \lambda_0 \Ndist(y_t \mid \mu_0, \sigma_0^2) 1_{(\ell = 0)} + \lambda_\ell \Ndist(y_t \mid \mu_\ell + f_{t,\ell}, \sigma_\ell^2) 1_{(\ell > 0)} } { \lambda_0 \Ndist(y_t \mid \mu_0, \sigma_0^2) + \sum_{j=1}^L \lambda_j \Ndist( y_t \mid \mu_j + f_{t,\ell}, \sigma_j^2 ) } \, ,
\end{align}
for $\ell = 0, 1, \ldots, L$, and $t=L+1, \ldots, T$, where in this context, $\Ndist(\cdot \mid \mu, \sigma^2)$ denotes a Gaussian density function with mean $\mu$ and variance $\sigma^2$.

The full Gibbs sampler for the GPMTD model then proceeds as follows:
\begin{enumerate}
	\item Draw $z_t$ from the discrete full conditional distribution given in (\ref{eq:zetaUpdate}) independently for $t = L+1, \ldots, T$.
	\item Calculate the current mixture allocation counts $\bm{n} = (n_0, n_1, \ldots, n_L)$ where $n_\ell = \sum_{t} 1_{(z_t=\ell)}$ and draw $\bm{\lambda}$ from the SBM-multinomial full conditional distribution outlined in Appendix \ref{sec:appendix_SBM}. A Dirichlet 
	prior for $\bm{\lambda}$ could also be trivially accommodated in this model, with this full conditional update corresponding to the conjugate model for multinomial data.
	\item Draw $\mu_0$ from the full conditional distribution $\Ndist \left(m_1^{(0)}, v_1^{(0)} \right)$ where \\ 
	$v_1^{(0)} = \left( (v_0^{(0)})^{-1} + n_0 / \sigma_0^2 \right)^{-1} $ and $m_1^{(0)} = v_1^{(0)}\left( m_0^{(0)}/v_0^{(0)} + \sum_{t:z_t=0}y_t / \sigma_0^2  \right) $.
	\item Draw $\sigma_0^2$ from the full conditional inverse-gamma distribution with shape $\left( \nu_\sigma^{(0)} + n_0 \right) / 2$ and scale $ \left( \nu_\sigma^{(0)} s_0^{(0)} + \sum_{t:z_t = 0} ( y_t - \mu_0 )^2 \right) / 2$.
	\item Perform the scan for $(\mu, \sigma^2, \bm{f}, \kappa, \psi)_\ell$ described in Appendix \ref{sec:appendix_mixcomp_updates}, independently for $\ell=1,\ldots,L$.
	\item Draw $\nu_\kappa$ and $\nu_\psi$ from their discrete full conditional distributions \\ $p(\nu_\kappa \mid \ldots) \propto \prod_{ \{ \ell > 0 : n_\ell > 0 \} } \left[ \IGdist(\kappa_\ell \mid \nu_\kappa / 2, \nu_\kappa \, \kappa_0 / 2) \right] \, 1_{(\nu_\kappa \in \mathcal{V}_\kappa)} $ and \\ $p(\nu_\psi \mid \ldots) \propto  \prod_{ \{ \ell > 0 : n_\ell > 0 \} } \left[ \IGdist(\psi_\ell \mid \nu_\psi / 2, \nu_\psi \, \psi_0 / 2) \right] \, 1_{(\nu_\psi \in \mathcal{V}_\psi)} $.
	\item Draw $\kappa_0$ and $\psi_0$ from their full conditional gamma distributions. In the former case, if we let $n^* = \sum_{\ell=1}^L 1_{(n_\ell > 0)}$ and $\tilde{\kappa} = \sum_{\{\ell > 0 : n_\ell > 0\}} \kappa_\ell^{-1}$, we have $p(\kappa_0 \mid \cdots) \propto \kappa_0^{a_\kappa + \nu_\kappa n^*/2 - 1} \exp \left[ -(b_\kappa + \nu_\kappa \tilde{\kappa}/2 ) \kappa_0 \right] \propto \Gammadist( \kappa_0 \mid a_\kappa + \nu_\kappa n^*/2, b_\kappa + \nu_\kappa \tilde{\kappa}/2) $. The full conditional distribution for $\psi_0$ is analogous.
\end{enumerate}

\subsection{Inference and forecasting}

Given posterior samples of model parameters fit through time $T$, it is straightforward to obtain a forecast distribution and other important quantities, including posterior uncertainty, for $y_{T+1}$. For each sample, one may calculate the first line of (\ref{eq:gpmtd1}) over a grid of $y_{T+1}$ values to estimate the one-step-ahead forecast distribution. Likewise, one may replace each distribution in (\ref{eq:gpmtd1}) with conditional means to obtain the forecast mean. This procedure extends to transition mean and density estimates for any fixed values of inputs $(y_{t-1}, \ldots, y_{t-L})$ by evaluating (\ref{eq:gpmtd1}) over a multidimensional grid of values for each posterior sample of model parameters.

Calculation of transition density and mean estimates requires values for each lag ($\{ y_{t-\ell} \}_{\ell = 1}^L$), regardless of inferences for $\bm{\lambda}$. However, one may be interested in these quantities conditional on a certain configuration of active lags. Suppose that inference for $\bm{\lambda}$ in a model fit using $L=3$ indicates that only the first two lags carry significant weight. One may specify a grid of values for the first two lags over which to evaluate (\ref{eq:gpmtd1}), substitute dummy or default values, such as the mean, for $y_{t-3}$, and examine the transition density or mean as a function of $y_{t-1}$ and $y_{t-2}$ only. We urge testing the resulting inferences for sensitivity to the default values used for inactive lags before making conclusions. For example, one could replace mean values for inactive lags with random values drawn uniformly across the range of $\{ y_t \}$.

Finally, one may make $K$-step-ahead forecasts by inductively simulating $(z, y)_{T+k}$ pairs, for $k=1,\ldots,K$, following the first two levels of (\ref{eq:hier}), for each posterior sample. The primary challenge here lies in the need to extend the $\{ \bm{f}_\ell \}$ Gaussian process realizations to include the $f_\ell(y_{T+k-\ell})$ that do not already exist, for which a naive computation approach involves repeatedly inverting a growing covariance matrix. When repeated for each posterior simulation, this results in a computational burden commensurate with MCMC. Given a current model state (i.e., full sample of all model parameters) the procedure to draw $f_\ell(y_{T+k-\ell})$ begins by calculating $c_{k} = \kappa_\ell \, \sigma_\ell^2$, and $(\bm{c}_k)_i = \kappa_\ell \, \sigma_\ell^2 \, \rho( y_{T+k-\ell}, x_i ; \nu, \psi_\ell)$ for all $x_i$ associated with the entries $f_{i,\ell}$. Then using the existing $\bm{f}_\ell$, draw a realization $f_\ell(y_{T+k-\ell}) \sim \Ndist \left( \bm{c}_k' (\bm{C}^{(\ell)})^{-1} \bm{f}_\ell ,\ c_k - \bm{c}_k' (\bm{C}^{(\ell)})^{-1} \bm{c}_k \right) $, where $\bm{C}^{(\ell)}$ is the existing covariance matrix for $\bm{f}_\ell$. Lastly, concatenate $\bm{C}^{(\ell)}$ with $c_k$ on the diagonal and $\bm{c}_k$ along an outer column and row, and concatenate $\bm{f}_\ell$ with the new draw from $f_\ell$. One can avoid re-calculating the new $(\bm{C}^{(\ell)})^{-1}$ from scratch by storing the previous inverse and using the inversion formula for partitioned matrices \citep[p.\ 201]{rasmussen2006gp}.

\section{Illustrations}
\label{sec:illustrations}

We demonstrate properties of the GPMTD model with two simulated and two real time series. The first simulation in Section \ref{sec:appdataLag2} highlights lag selection and nonlinear dynamics. The second simulation in Section \ref{sec:appdata2d} explores the model's fitness for approximating higher order dynamics in a time-delay embedding context. We then apply the GPMTD to a noisy time series known for nonlinear and non-Gaussian transitions in Section \ref{sec:oldFaithful}, and finally to a time series for which we anticipate a certain lag dependence structure in Section \ref{sec:pink}.


Each of the following analyses included at least three MCMC runs with chains initialized at default values (i.e., independent standard normal mixture components, uniform $\lambda$, and all observations allocated to the intercept). A Metropolis adaptation phase was followed by 5,000 burn-in iterations. A final run of 10,000 iterations was thinned to 2,000 inference samples (1,000 were used for some two-dimensional plots), which are reported for one chain. Unless otherwise reported, inferences for functionals of (\ref{eq:gpmtd1}) with respect to fewer than $L$ lags were obtained by inserting default mean values for inactive lags, which could be identified, for example, as $\{ x_{t,\ell} : \E(\lambda_\ell \mid \{ y_t \} ) < c_\lambda \}$ for some small positive value $c_\lambda$ (such as 0.01).

\subsection{Simulated data: single lag}
\label{sec:appdataLag2}

We first demonstrate lag selection with a nonlinear time series simulated from a classical model for population dynamics \citep{ricker1954}. The series was generated from
\begin{align}
	\label{eq:appDatalag2gen}
	y_t = y_{t-2} \exp(2.6 - y_{t-2}) + \epsilon_t \, , \quad \epsilon_t \simiid \Ndist(0, (0.09)^2) \, ,
\end{align}
featuring first-order nonlinear dynamics, and specifically adapted to be a function of the second lag only. We fit the GPMTD model to the real-valued time series with $L=5$ and $T=105$ (so that 100 observations contribute to the likelihood). All three MCMC chains converge to the same region of the parameter space, although one earlier run showed a posterior mode with observations allocated to the fourth lag, whose marginal relationship to the current observation resembles that of the second lag.

\begin{figure}[b!]
    \includegraphics[width=3.0in]{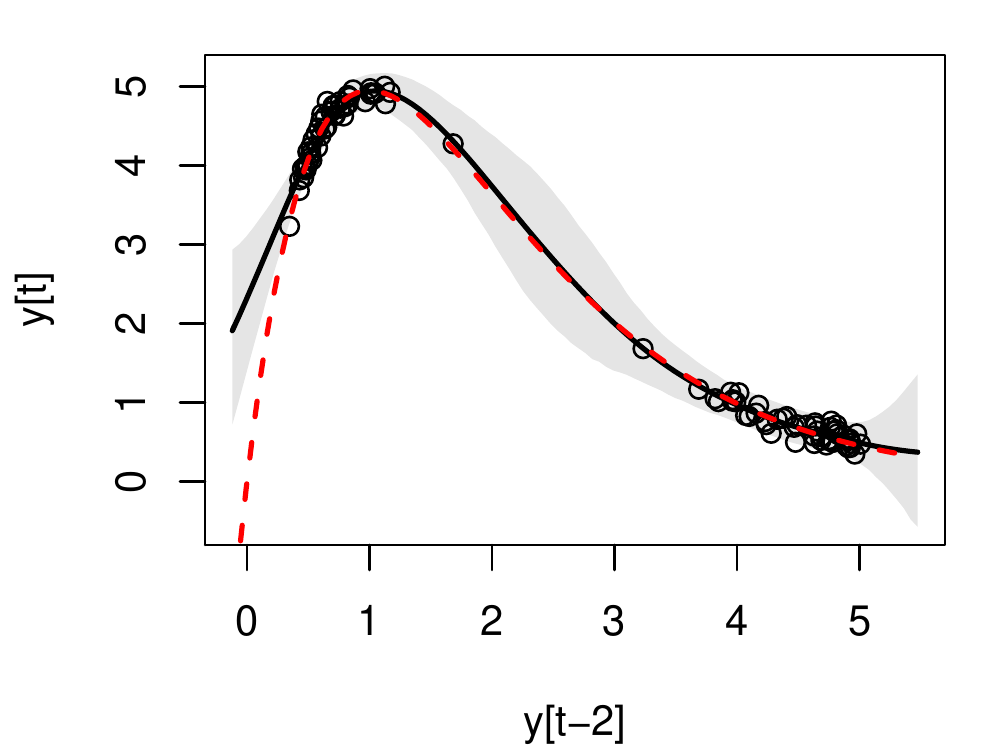}
    
	\caption{GPMTD fit to the single-lag dynamical simulation with noise. The solid black curve depicts the model estimate of the overall transition mean as a function of the second lag only, together with a 95\% credible interval shaded in gray. The true transition mean function is given by the dashed curve. All observed two-step transitions are included as points.}
	\label{fig:appdata_lag2only_transMean}
\end{figure}

In this example, the model decisively recovers the true structure. Inferences strongly favor using one lag, with the 0.025 posterior sample quantile of $\lambda_2$ being greater than 0.99. The estimated transition mean as a function of $y_{t-2}$ (holding other lags fixed), with 95\% pointwise credible intervals, is shown in Figure \ref{fig:appdata_lag2only_transMean} together with the data and true transition mean function. The dynamics are successfully recovered within the range of observed transitions, except on the far left, where the estimated curve tends back toward the component level $\mu_2$ (which has posterior mean around 0.9, and standard deviation 2.4) in a smooth manner. This likely stems from stationarity of the covariance function and bias from the default prior on the component-specific observation variance $\sigma_2^2$. Also, one could argue that the stationary covariance function does not allow sufficient uncertainty in the central region ($y_{t-2} \in (2,3)$) with no observed transitions.

\subsection{Simulated data: time-delay embedding}
\label{sec:appdata2d}

Our second simulation example explores the GPMTD model's fitness for approximating higher-order dynamics. We do so with an example of statistical state-space reconstruction via time-delay embedding, which attempts to reconstruct a multidimensional attractor using lags from a single time series. The modeling objective for this example is to infer a suitable embedding dimension and estimate the corresponding transition map. 

\begin{figure}[b!]
  \begin{center}
    \includegraphics[width=5in, trim = 0 0 0 10, clip]{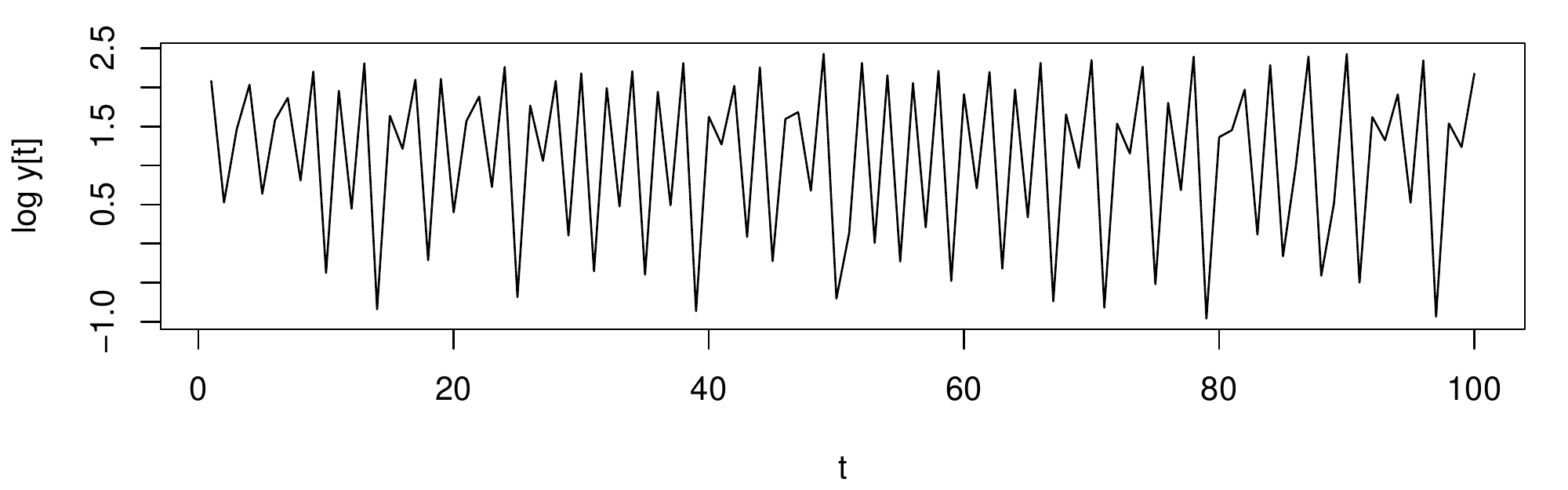}
  \caption{Trace of 100 steps of the log-transformed $y_t$ series from the simulated deterministic nonlinear system.}
  \label{fig:2dsys_timeseries}
  \end{center}
\end{figure}

\begin{figure}[t!]
  \begin{center}
      \includegraphics[width=2.75in]{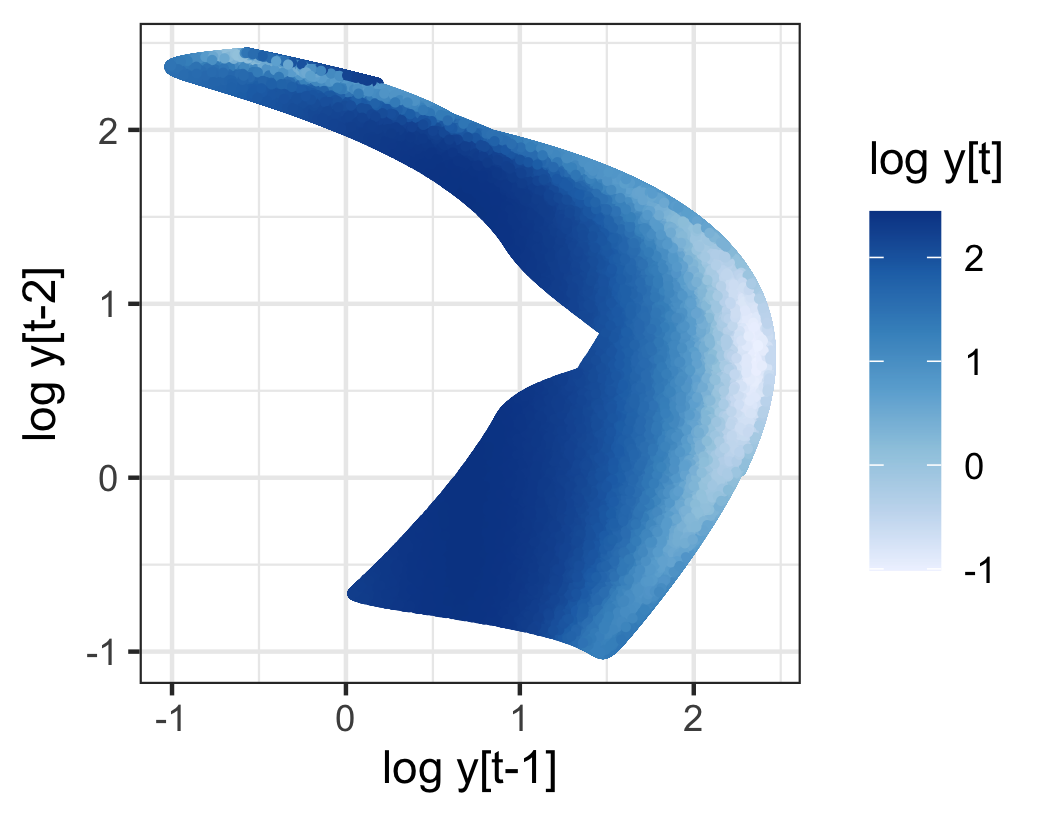}
    \caption{Transition surface for the time-delay embedding of $\log(y)$ from the nonlinear deterministic system (\ref{eq:sim2dsys}), as represented by simulated time steps. All points lie exactly on the surface. The plot was generated with {\it ggplot2} \citep{ggplot2}.}
    \label{fig:2dsys_truelogtde}
  \end{center}
\end{figure}

We first simulated a long time series (with sufficient burn-in) from the following two-dimensional deterministic system used to represent predator-prey dynamics with interaction \citep{basson1997ricker},
\begin{align}
  \label{eq:sim2dsys}
  y_t &= y_{t-1} \exp ( r - a y_{t-1} - b z_{t-1} ) \, ,  \\
	z_t &= z_{t-1} \exp ( r - a z_{t-1} + b y_{t-1} ) \, , \nonumber
\end{align}
using $r=2.75$, $a=0.5$, and $b=0.07$. In this case, substitution yields an analytical expression for a time-delay embedding of this system in two lags using either the $\{ y_t \}$ or $\{ z_t \}$ series alone. The resulting transition surface for the $\{ y_t \}$ series is more regular when we consider the dynamics on the $\log(y)$ scale, a natural transformation in this scenario. 
A trace for 100 successive values of $\log(y_t)$ is shown in Figure \ref{fig:2dsys_timeseries}. Figure \ref{fig:2dsys_truelogtde} shows the time-delay embedding transition surface for $\log(y)$, one of the inferential targets in this example. Note that the continuous surface pivots sharply along the right and upper border of observed lag combinations, with points falling on both sides of a steep and narrow trench, beyond which the surface exhibits super-exponential growth.



It is immediately apparent that the GPMTD model is inadequate to fully capture this non-additive, intricate function of two lags. If the model admitted general functions of two inputs, or at least interactions, one could enforce near determinism with the priors on component-specific variances $\{ \sigma_\ell^2 \}$. Unless the modeler is confident that only one lag is active, we discourage this practice with the GPMTD for two reasons. The first is that the model will attempt to interpolate apparent noise in each one-dimensional projection, which occurs in this example. Second, the mixture of densities defining the model will produce multiple highly separated modes for most combinations of lag values. For these reasons, we forego pursuing a high-fidelity estimate of the transition surface with the GPMTD, allowing for observation noise to instead smooth over finer features of the surface.

As before, we fit the GPMTD model with default priors and initial values to a $\{ \log(y_t) \}$ series of length $T=105$ and $T=505$ using a lag horizon of $L=5$. All three chains converge to similar log-likelihood values and lag configuration for the shorter time series. Two of the three chains likewise converge for the longer series, while one chain remains stuck at a mode with significantly lower log-likelihood.

\begin{figure}[t]
    \begin{center}
        \includegraphics[height=2.4in]{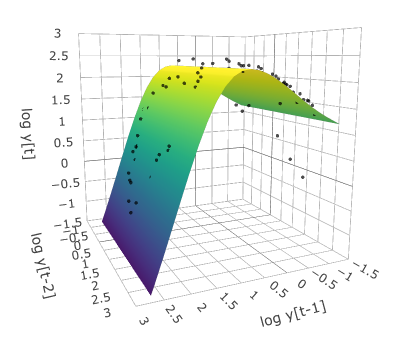}
        \includegraphics[height=2.4in]{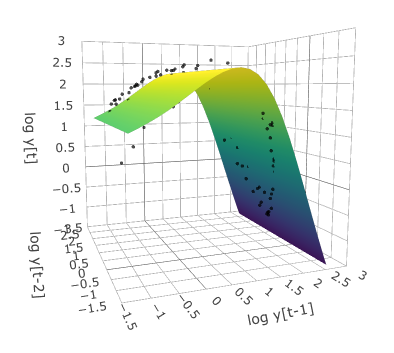}
      \caption{GPMTD model fit ($T=105$ and $L=5$) to the time-delay embedding of $\log(y)$ simulated from the nonlinear deterministic system. Plots include the posterior mean estimate for the transition surface and observed transitions as points. All multidimensional plots were generated with {\it Plotly} \citep{plotly}.}
      \label{fig:appdata2d_log_n100_meanSurf}
    \end{center}
\end{figure}

The model fit to the shorter time series produces mixed results. It selects lag 1 only (with a 0.025 posterior sample quantile above 0.99), which appears reasonable given the sample size and the fit depicted in Figure \ref{fig:appdata2d_log_n100_meanSurf}. The model fails to capture only a few points in the border trench along $\log(y_{t-1}) \in (-0.5, 0.5)$, $\log(y_{t-2}) \approx 2.3$. Because these observations are not allocated to another mixture component and treated as outliers, the component-specific standard deviation (effectively the global error standard deviation since $\lambda_1 \approx 1$) is estimated high at 0.4.

\begin{figure}[t!]
  \begin{center}
      \includegraphics[width=2.75in]{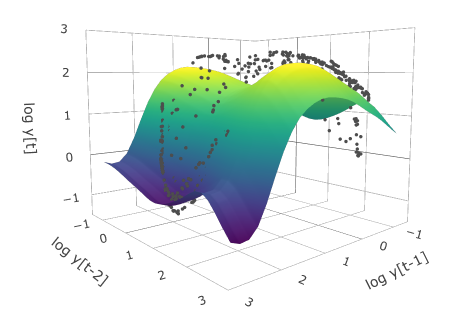}
      \includegraphics[width=2.75in]{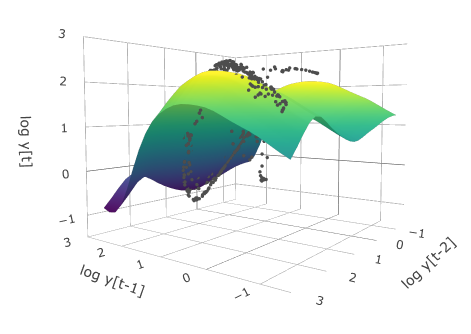} \\
      \includegraphics[width=2.75in]{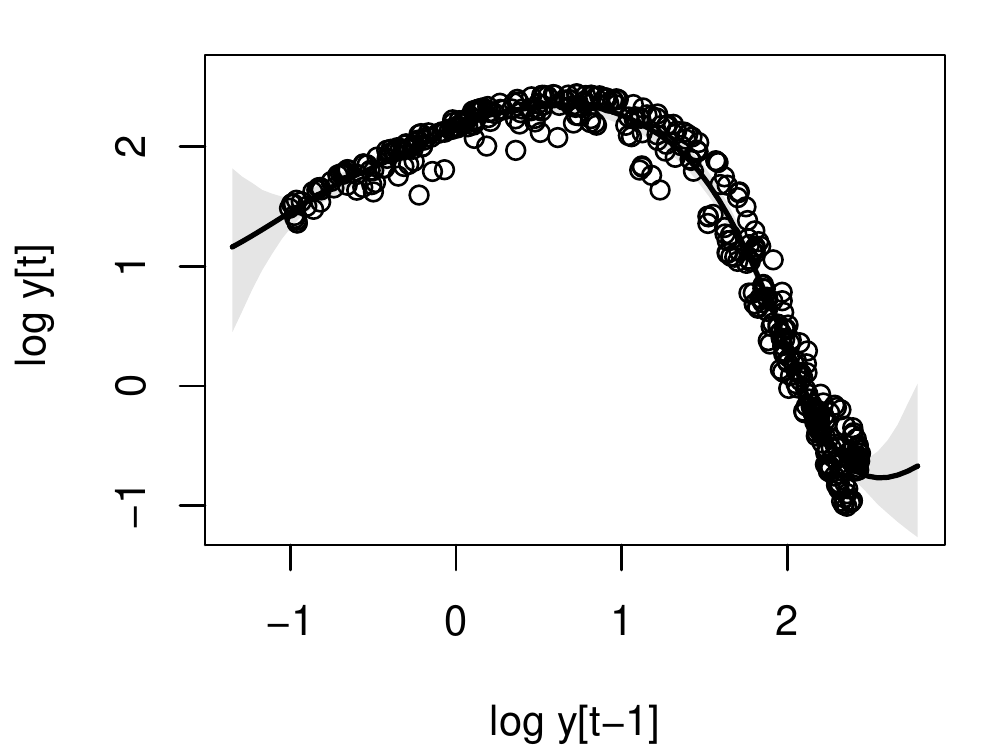}
      \includegraphics[width=2.75in]{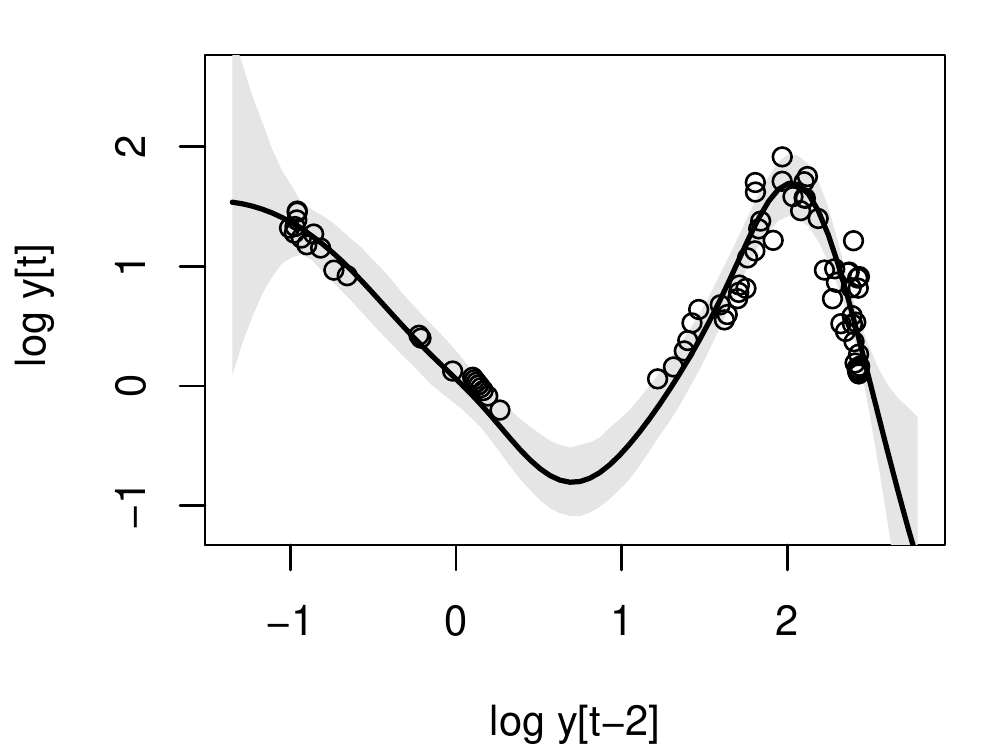}
    \caption{GPMTD model fit ($T=505$ and $L=5$) to the time-delay embedding of $\log(y)$ simulated from the nonlinear deterministic system. Posterior mean estimate for the transition surface (top) and lag-specific $f_1$ and $f_2$ functions (with pointwise 95\% intervals, bottom). Data values are included as points. In the lower plots, points are included with a lag if allocated to that lag (with posterior probability greater than 0.5).}
    \label{fig:appdata2d_log_n500_meanSurf}
  \end{center}
\end{figure}

The model fit to the longer time series provides a surprisingly robust approximation, considering the level of model mis-specification. Two lags are selected, with $\lambda_1$ and $\lambda_2$ receiving a 0.78, 0.22 split in posterior mean (both 95\% intervals have approximate length 0.12). The posterior mean estimate of the transition surface is given in Figure \ref{fig:appdata2d_log_n500_meanSurf}, together with marginal estimates of $f_1$ and $f_2$ and their assigned observations (classified if the observations are assigned to the corresponding lag with at least 0.5 posterior probability). The most obvious omission in the estimated surface is the outer wall or border. This is expected, as the lower trench is not clearly identified in one dimension. Assuming noisy observations, $f_1$ and $f_2$ fit the corresponding one-dimensional projections well, while the overall estimated transition surface appears attenuated, a result of the global mixture. Similarly, transition density estimates (not shown) for lag values along the two shoulders and central dip of the surface are bimodal with small component variances. The second-lag component successfully captures the ``outliers'' near $\log(y_{t-1}) \approx -0.75$, $\log(y_{t-2}) \approx 2.3$, producing an appropriate mixture transition density in this region.

Overall, we caution that despite its ability to produce GAM-like estimates for transition surfaces, the lag-dependent error structure of the GPMTD model is not suited to applications for high-order dynamics, unless functions of the correct order are explicitly included. The model is better poised to estimate possibly nonlinear, lag-dependent transition densities in the presence of noise, as with the two examples that follow.

\subsection{Old Faithful data}
\label{sec:oldFaithful}

\begin{figure}[b!]
  \includegraphics[scale=0.65, trim = 0 10 0 40, clip]{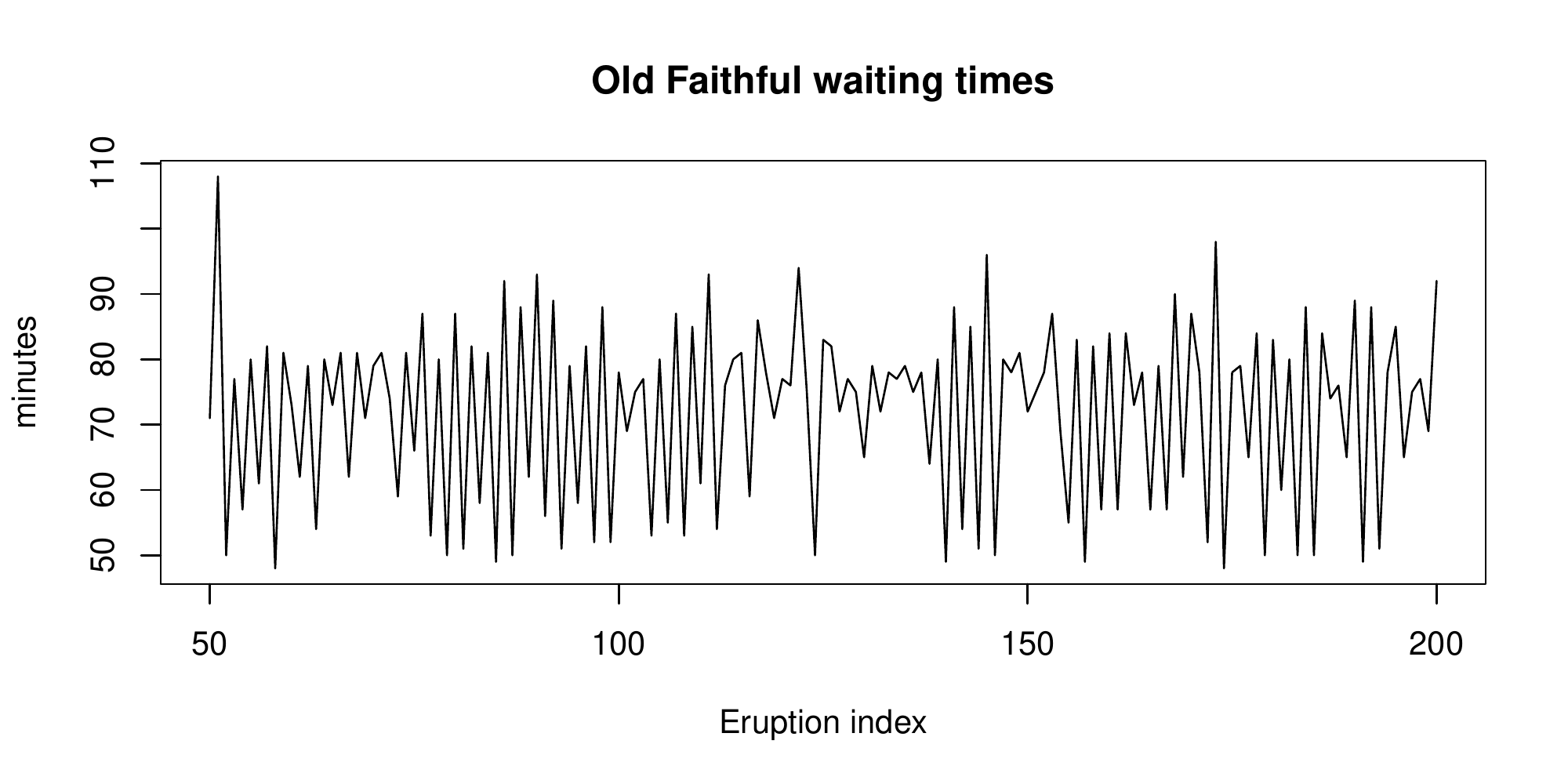}

\caption{Trace of 150 consecutive Old Faithful eruption waiting times in minutes (top). This window of the middle half of the time series typifies the data, with exception of the run of long waiting times between index 120 and 140. 
}
\label{fig:OldFaithful_ts}
\end{figure}

\begin{figure}[p]
    \includegraphics[scale=0.7]{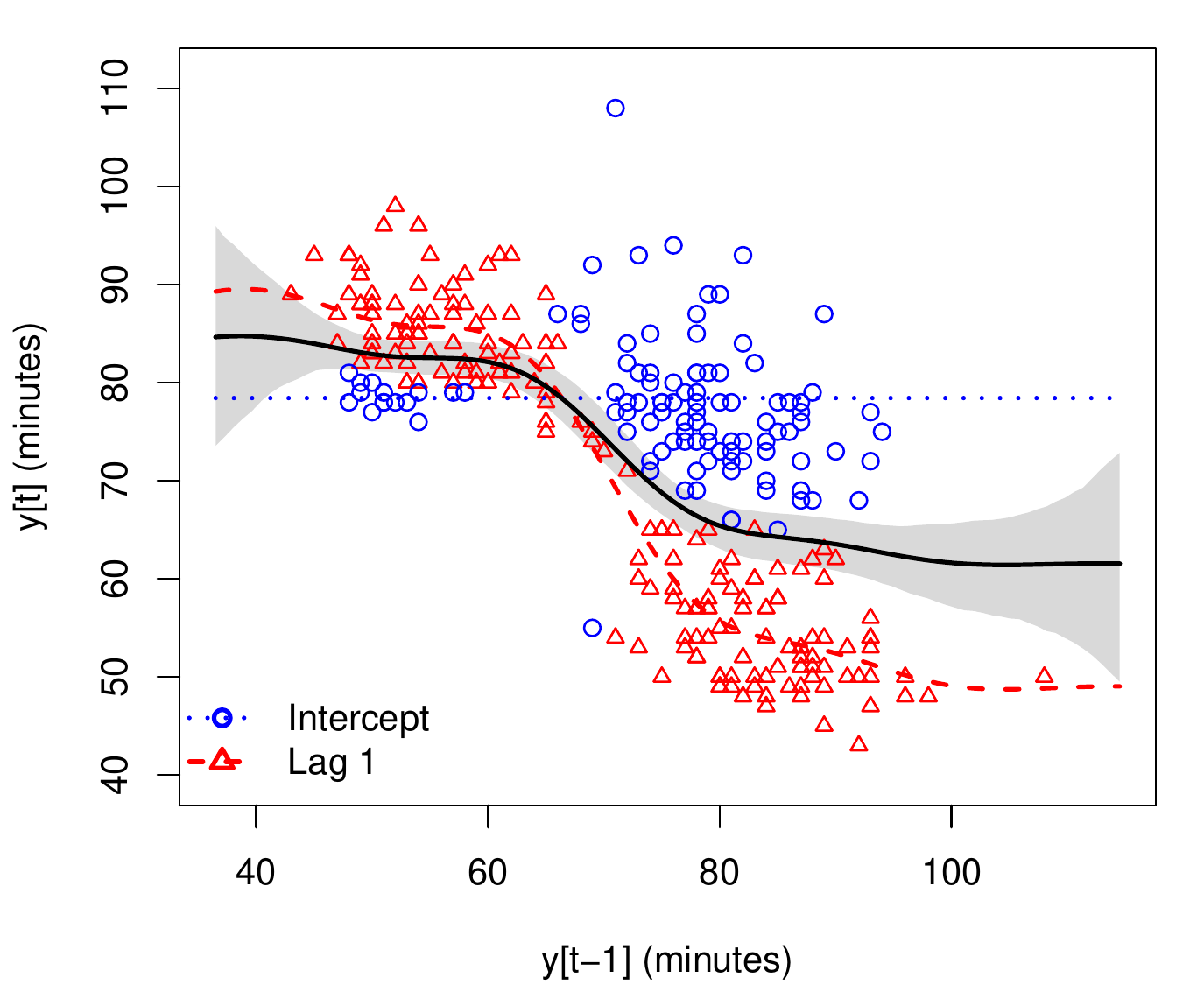}
    
	\caption{Single-step transition scatter plot with component-specific inferences from the GPMTD fit to Old Faithful waiting times. Circles indicate membership in the intercept mixture component (with posterior probability greater than 0.5), and triangles indicate the same for the first lag mixture component. Dashed curves report the posterior mean for the respective component means. The solid curve depicts the model estimate of the overall transition mean, together with a 95\% credible interval shaded.}
	\label{fig:OldFaithful_transMean}
\end{figure}

Our first illustration of the GPMTD with real data highlights the model's capability for capturing heterogeneous lag dependence with the well-known series of inter-eruption waiting times of the Old Faithful geyser in Yellowstone National Park, U.S.A. The time series has attracted attention, both for illustration and analysis from chaos \citep{nicholl1994} and statistical \citep{azzalini1990} perspectives, partly due to nonlinear as well as non-Gaussian dynamics. We revisit Old Faithful using the traditional data set reported in \citet{azzalini1990}, consisting of 299 consecutive pairs of eruption durations and waiting times between August 1 and 15, 1985. Figure \ref{fig:OldFaithful_ts} shows a trace of eruption waiting times in minutes. 
Despite high noise levels, dependence on at least one lag is clearly discernible among the raw values depicted as points in Figure \ref{fig:OldFaithful_transMean}. The relationship between consecutive waiting times appears mostly consistent across values of the second lag, but a trend may exist (not shown). 
The GPMTD model is unlikely to detect higher-order dynamics. 
We do, however, expect the model to capture the apparent nonlinear and non-Gaussian dependence on the first lag.

GPMTD model runs with $L=5$ and $L=10$ all indicate dependence on one lag. All chains converge to the same region in the parameter space, with exception of one run with $L=5$ that switched to an allocation with some observations assigned to the second lag. We report results from one of the runs with $L=10$. Mixture weights $\lambda_0$ and $\lambda_1$ dominate, accounting for more than 99\% of the allocation in the posterior mean of $\bm{\lambda}$. Point estimates and 95\% credible intervals for each $\lambda_\ell$ are reported in Table \ref{tab:OldFaithful1985_postlambda}. The intercept carries significant weight in order to provide bimodality in the transition distribution, while the first lag component captures nonlinear dependence.

The model's use of both intercept and first lag is apparent in Figure \ref{fig:OldFaithful_transMean}, which shows a superimposed scatter plot on the first lag indicating inferred mixture allocation. Circles are assigned to the intercept component with posterior probability greater than 0.5, and triangles are likewise assigned to the first lag. The corresponding dashed curves give posterior mean inferences for the respective component means. The solid curve depicts the pointwise estimate of the transition mean, together with a 95\% credible interval shaded. The transition mean is less useful for lagged values above 70 minutes, where it begins to straddle the bimodal transition density.

\begin{table*}[t]
	\centering
	
    \caption{Posterior summary for $\lambda_\ell, \ \ell = 0, \ldots, 10$ in the GPMTD analysis of Old Faithful waiting times. 
    Lag $\ell = 0$ refers to the intercept.}
	\label{tab:OldFaithful1985_postlambda}
	\small
		\begin{tabular}{rrr}
			\toprule
			$\ell$ & Mean & 95\% Interval \\
			\midrule
			0 & 0.428 & (0.332, 0.512) \\
			1 & 0.571 & (0.486, 0.666) \\
			2 & $<$0.001 & ($<$0.001, 0.002) \\
            3 & $<$0.001 & ($<$0.001, 0.001) \\
            4-10 & $<$0.001 & ($<$0.001, $<$0.001) \\
            \bottomrule
		\end{tabular} 	
\end{table*}

\begin{figure}[p]
    \includegraphics[scale=0.57, trim=5 0 10 0, clip]{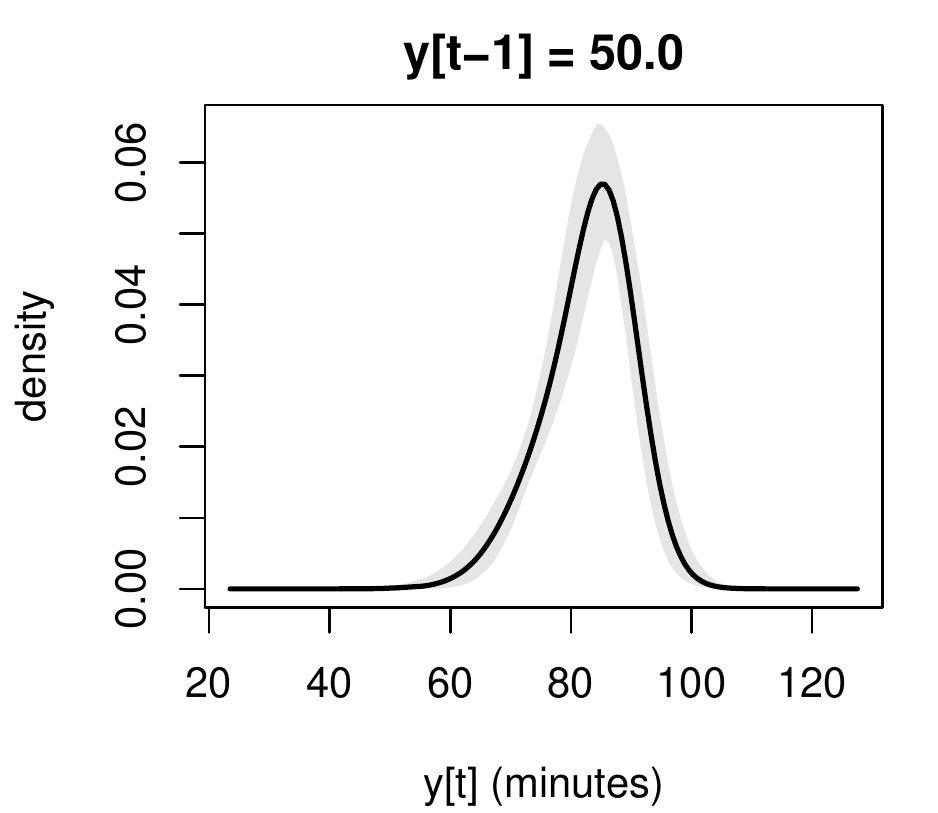}
    \hspace{-5pt}
    \includegraphics[scale=0.57, trim=30 0 10 0, clip]{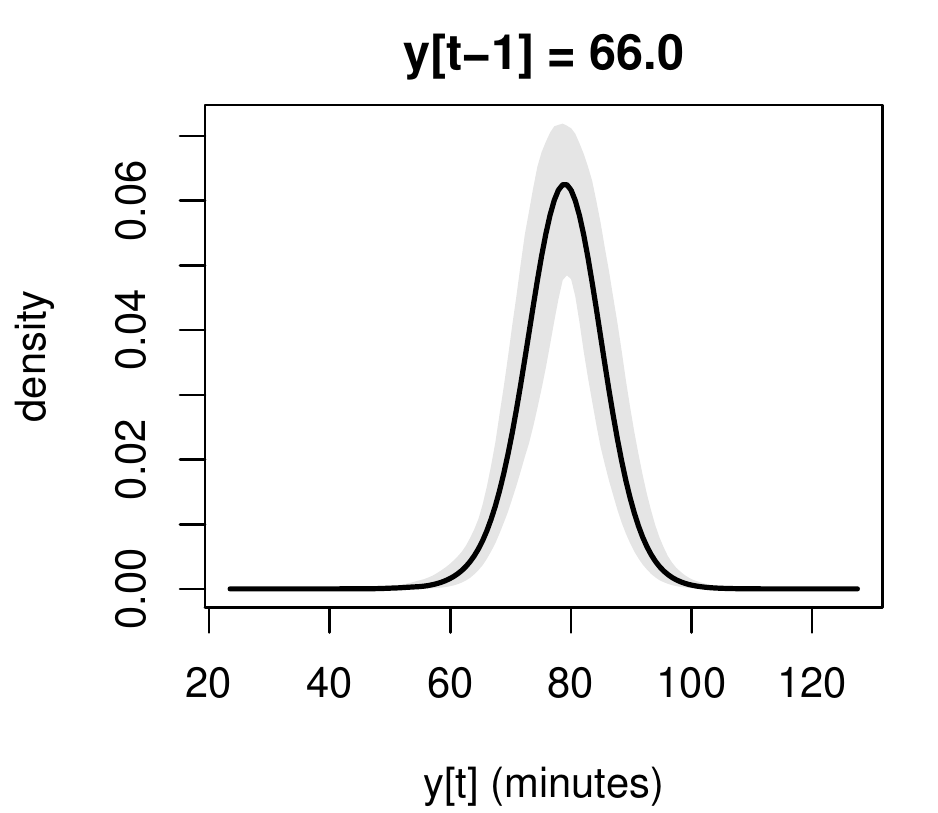}
    \hspace{-5pt}
    \includegraphics[scale=0.57, trim=30 0 10 0, clip]{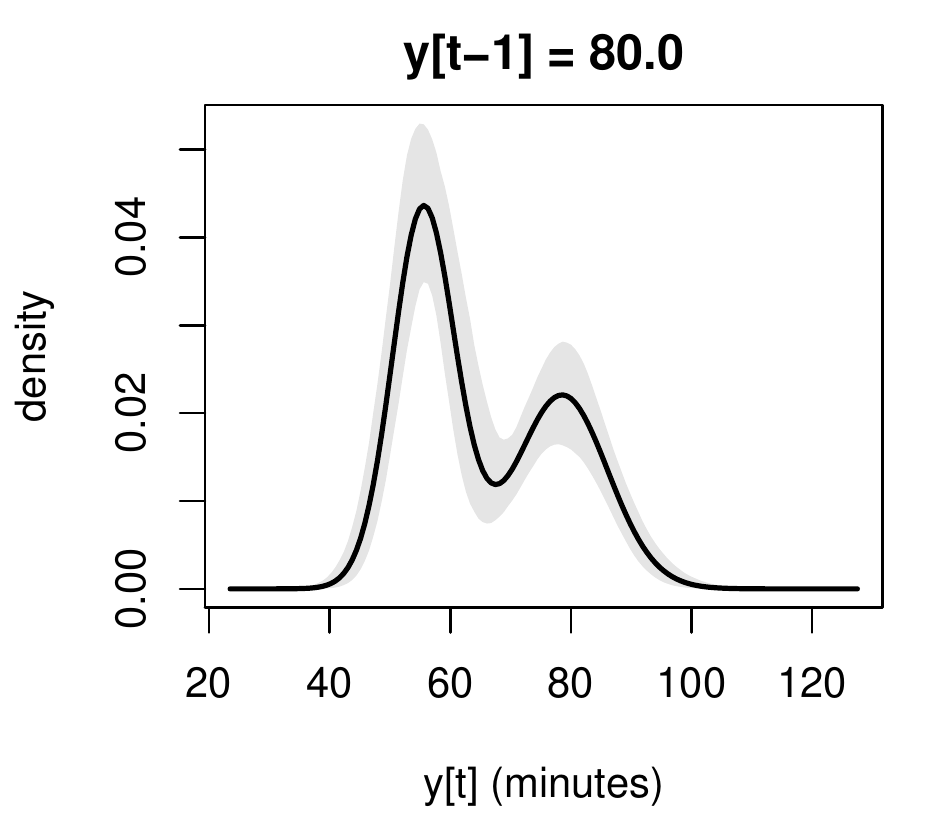}
    
	\caption{GPMTD transition density estimates for Old Faithful waiting times at three fixed values of the first lag: $y_{t-1}=50$, $y_{t-1}=66$, and $y_{t-1}=80$ minutes. The solid line indicates the pointwise posterior mean and gray shading indicates 95\% intervals.}
	\label{fig:OldFaithful_densities}
\end{figure}

Figure \ref{fig:OldFaithful_densities} highlights the model's flexibility in approximating transition densities, summarizing posterior inferences of the density for three values of the first lag $y_{t-1} \in \{ 50, 66, 80 \}$. These estimates are generated almost exclusively from a two-component mixture, and should therefore be treated as rough approximations. For example, because the more concentrated density associated with lag 1 is located above the mean of the wide intercept density at $y_{t-1}=50$, the model yields a left-skewed density for $y_{t-1} = 50$, while one could argue from Figure \ref{fig:OldFaithful_transMean} that the transition density at this lag should exhibit right skew. The means of mixture components $\ell=0$ and $1$ intersect near $y_{t-1}=66$, appropriately resulting in a scale mixture of normal distributions for the transition. At $y_{t-1}=80$, the mixture captures the obvious bimodality.

Nonlinear, heterogeneous, and noisy lag dependence makes the Old Faithful time series a unique candidate for illustrating both strengths of the GPMTD model. When the model employs mixing for both transition density approximation {\em and} nonlinear transition surface estimation simultaneously, we entreat practitioners to carefully scrutinize and validate inferences. For example, because the mixing weights are global, they may not be optimized for the transition density specifically at $y_{t-1}=80$ in the Old Faithful model. While the parsimonious representation (\ref{eq:gpmtd1}) has such limitations, it is quite flexible relative to the mixtures of linear autoregressive models in the literature, efficiently capturing nonlinear and non-Gaussian dynamics.


\subsection{Pink salmon data}
\label{sec:pink}

\begin{figure}[b]
	\centering
	
	\begin{tabular}{cccc} 
		\multicolumn{4}{c}{\includegraphics[scale=0.55]{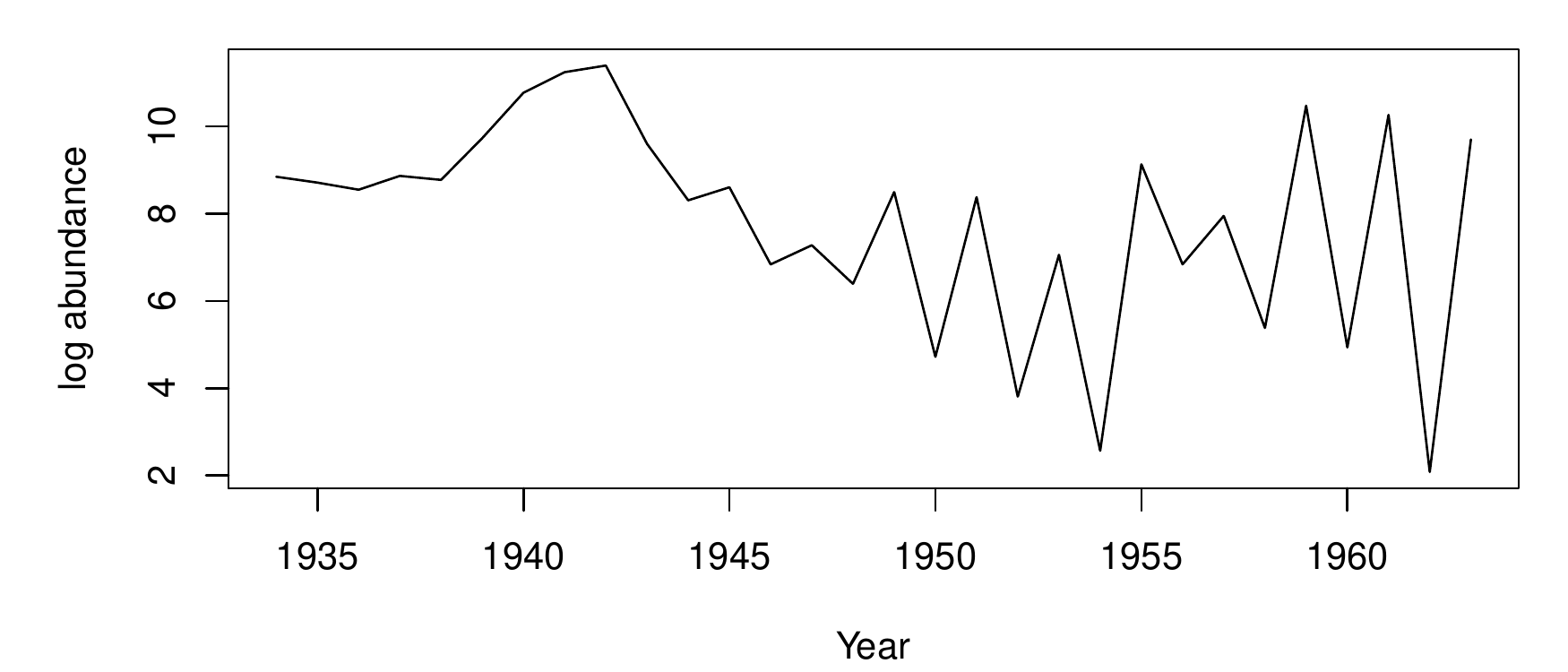}}
	\end{tabular}
	
	\caption{Trace of the natural logarithm of pink salmon abundance from 1934 to 1963.} 
	\label{fig:pinksalmon_timeseries}
\end{figure}

We next investigate a time series of annual pink salmon abundance (escapement) in Alaska, U.S.A.\ 
from 1934 to 1963 \citep{PinkSalmonData}. Population dynamics for pink salmon provide an opportunity to test selection in the GPMTD model since pink salmon have a strict two-year life cycle \citep{heard1991pinksalmon}. Thus, we expect even lags to have the most influence in predicting the current year's population. The trace of the natural logarithm of abundance in Figure \ref{fig:pinksalmon_timeseries} suggests a comprehensive analysis might appropriately include non-stationarity with long-term trends, which we forego in favor of a simple demonstration. 
Repeated interventions for the struggling even-year salmon population throughout the 1950s culminated in a population transfer in 1964 that bisects the complete time series and restricts us to the first segment \citep{PinkSalmonReport}. Nevertheless, lag scatter plots (not shown) suggest that we should be able to detect lag dependence structure, even with as few as 30 observations.

\begin{figure}[t!]
    \includegraphics[scale=0.75]{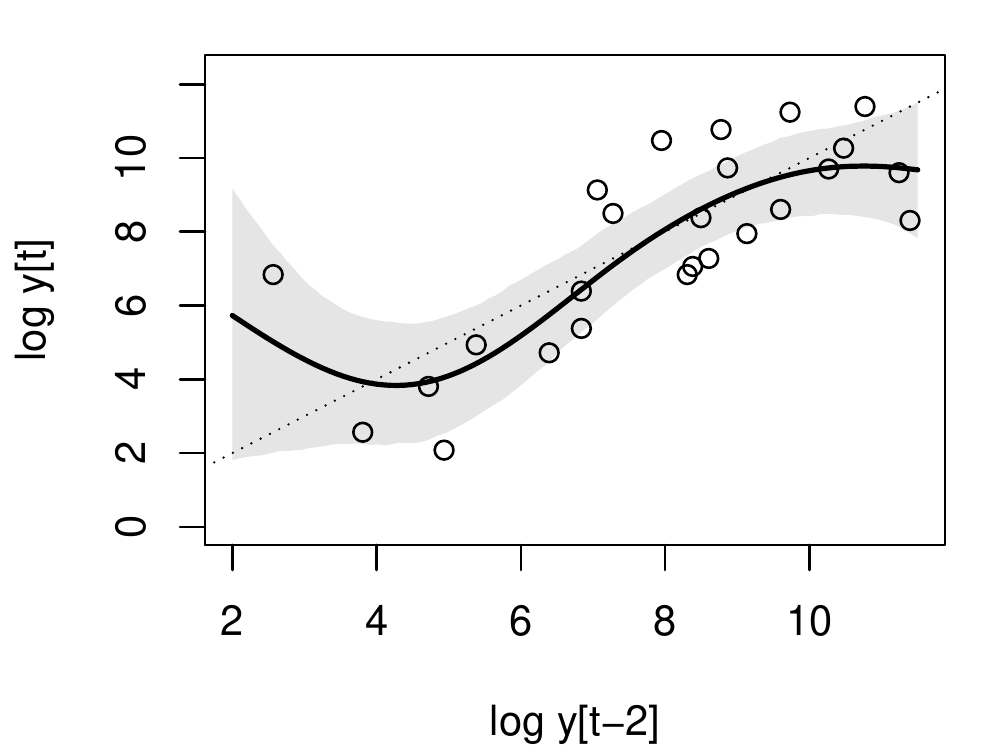}
    
	\caption{GPMTD fit to the logarithm of annual pink salmon escapement, with a scatter plot of all two-step transitions. The solid black curve gives the overall transition mean, together with a 95\% credible interval shaded in gray. The reference line has unit slope and passes through the origin.}
	\label{fig:Pink_transMean}
\end{figure}

We fit the GPMTD with up to $L=5$ lags to the logarithm of annual escapement using the same default prior, initialization, and MCMC sampling employed for other analyses. All chains converge to the same estimated posterior distributions. As expected, the model clearly identifies the second lag as dominant, as $\lambda_2$ has a posterior mean of 0.975 with a 95\% equal-tailed interval of (0.683, 0.999). Lags 1 and 4 have the next highest upper (0.975) quantiles at 0.095 and 0.046, respectively. The estimated transition mean function with pointwise 95\% intervals is given for the second lag (fixing other lags) in Figure \ref{fig:Pink_transMean}. The diagonal dotted line has a unit slope dividing regions of population increase and decrease. Although the interval seldom leaves this line, population decline is readily apparent, particularly with the even-year population (in Figure \ref{fig:pinksalmon_timeseries}), which experienced repeated interventions \citep{PinkSalmonReport}. Although inclusion of covariates or explicitly modeling interventions may increase signal resolution, the expected dependence structure is manifest in the raw series, and detected by the GPMTD model.

\section{Discussion}
\label{sec:discussion}

The model proposed in this article, building on a Bayesian framework for the continuous-state GMTD model, contributes two helpful extensions: 1) nonlinear transition dynamics, and 2) model-based order and lag selection. Although the original GMTD model accommodates non-Gaussian transition distributions and heterogeneity with a mixture, we further allow the mixture kernels to exhibit nonlinear lag dependence. Thus the Gaussian process mixture transition distribution model can be considered a parsimonious, semiparametric model for nonlinear transition density approximation. With care, the model can be further used to identify low-noise nonlinear dynamics in one lag.

We note the potential for confounding when the mixture model is used for both flexibility in density estimation and lag selection. For example, components (lags) could activate to add modes to a transition density when no dependence on the associated lag exists. The SBM prior affords the intercept priority in adding such flexibility. For example, the intercept component is instrumental for the Old Faithful example in that it provides a vehicle both for bimodality (when $y_{t-1} > 70$ minutes) and a pair of outliers (at $y_{t-1} \approx 70$ minutes). We urge practitioners to examine inferences for the mean transition function of all active lags. A flat transition function would indicate that the component contributes exclusively to density flexibility.

If certain characteristics of the transition distribution systematically associate with a certain lag, it is more appropriate to accommodate them within the corresponding mixture component. Adding flexibility to the mixture component distributions could further help disentangle the objectives of transition density estimation through mixtures and lag selection. For example, if the mixture is used primarily for lag selection, one could implement parametric extensions of Gaussian component transition densities \citep{hansen1994ARdensity} to allow for long-tails and/or skew without sacrificing parsimony or computational convenience.

The GPMTD model is inherently Markovian, directly modeling a probability distribution governing transitions. It consequently most naturally resides in the class of time-series models for dynamical (rather than measurement) error. It can nevertheless be extended, in a state-space framework or otherwise, to include other features common in time series, such as covariate dependence, trends, and periodic fluctuations. The most straightforward way to incorporate covariates 
is through additional mixture components dedicated to the exogenous variables. As this breaks the natural ordering of mixture components, one would need to reconsider the prior for $\bm{\lambda}$. Incorporating trends, periodicity, and covariates outside the MTD structure presents more of a challenge, as these would most naturally fit into a linear superposition with a latent GPMTD process. Estimation of GPMTD parameters would be no more complicated in such a model. However, updates for parameters governing external structures would necessitate re-evaluation of the GPMTD component-mean functions $\{f_\ell\}$ at each iteration of MCMC (or optimization), potentially creating a heavy computational burden.

The mixture autoregressive (MAR) model of \citet{wong2000mixture}, consisting of a finite mixture of Gaussian AR models of potentially varying order, is considered to be the sequel to the GMTD in the literature. While the MAR model indeed contains the GMTD as a special case, it does not generalize the linear transition mean, and perhaps more importantly, diverges from the parsimonious and interpretable representation as a mixture of low-order transition distributions. We have proposed and demonstrated a model that preserves these useful and distinguishing characteristics of the MTD and GMTD models. When data complexity demand richer models, the MAR and related mixtures of linear autoregressive models, such as mixtures-of-experts \citep{jordan1994hme, peng1996bayesmoe, carvalho2005, carvalho2006modeling} and nonparametric mixtures \citep{dilucca2013,antoniano2016,deyoreo2017,kalli2018npbvar} can provide added flexibility.


\section*{Data availability statement}
Simulated data referenced in Sections \ref{sec:appdataLag2} and \ref{sec:appdata2d} are provided in the supplementary materials. The Old Faithful data are published in \citet{azzalini1990} and available as the {\tt geyser} data set in the MASS package in R. The pink salmon data were obtained through \citet{PinkSalmonData}, reside in the public domain, and are provided in the supplementary materials.

\section*{Acknowledgements}
This research was supported in part by the National Science Foundation under award SES 1631963. The authors thank Stephan Munch for several helpful conversations, including motivation and materials for time-delay embedding applications. Chris Archibald, Candace Berrett, Gilbert Fellingham, Stephen Jones, and Richard Warr provided helpful editorial comments.

\begin{center}
{\large\bf SUPPLEMENTARY MATERIAL}
\end{center}

\begin{description}

  


 
  \item[GPMTD examples:] Directory containing code and data necessary for fitting and post-processing GPMTD examples from Section \ref{sec:illustrations} with the {\it GPMTD Julia} package (available at
  {\tt https://github.com/mheiner/GPMTD.jl.git}). The file README.md contains descriptions and instructions. (The GPMTD\_examples directory can be downloaded from {\tt https://github.com/mheiner/MTD\_examples.git})

\end{description}

\bibliographystyle{jasa3.bst}
\bibliography{Bibliography}


\appendix

\newpage

\section{Setup for mixture component updates} 
\label{sec:appendix_mixcomp_updates}

Conditional on the configuration variables $\{ z_t \}$ and covariance hyperparameters $\nu_\kappa$, $\kappa_0$, $\nu_\psi$, $\psi_0$, we have $L$ independent block-conditional updates for parameters in the (non-intercept) mixture components. To simplify notation, assume without loss of generality that we are working with component $\ell$, so that we can drop the $\ell$ index on each parameter. Let $n_\ell$ count the cardinality of $\{ t : z_t = \ell \}$ and partition $\bm{f}$ into $\bm{f}^{i}$ and $\bm{f}^o$, indexed by $z_t = \ell$ and $z_t \ne \ell$, respectively. The joint full conditional density for this component is
\begin{align}
	\label{eq:mixcompFCall}
	p(\mu, \sigma^2, \bm{f}, \kappa, \psi & \mid \{ z_t \}, \nu_\kappa, \kappa_0, \nu_\psi, \psi_0, \{ y_t \}) \propto \prod_{t:z_t = \ell}\left[ \Ndist(y_t \mid \mu + f_{t,\ell}, \sigma^2) \right] \times \nonumber \\
	& \Ndist(\mu \mid m_0, v_0) \ \IGdist( \sigma^2 \mid \nu_\sigma/2, \nu_\sigma s_0 / 2 ) \ \Ndist\left( \bm{f} \mid \bm{0}, \kappa \sigma^2 \bm{R}(\psi) \right) \times \\
	& \IGdist(\kappa \mid \nu_\kappa / 2, \nu_\kappa \kappa_0 / 2) \ \IGdist(\psi \mid \nu_\psi / 2, \nu_\psi \psi_0 / 2) \, , \nonumber
\end{align}
where the normal density for $\bm{f}$ has dimension $T-L$ and correlation matrix $\bm{R}(\psi)$ which can also be partitioned into active and inactive parts $\bm{R}^{ii}$, $\bm{R}^{oo}$, $\bm{R}^{io}$, and $\bm{R}^{oi} = (\bm{R}^{io})'$. We begin by marginalizing $\bm{f}$ out of (\ref{eq:mixcompFCall}), resulting in a $n_\ell$-variate Gaussian density for the vector $\bm{y}^i$ containing $\{ y_t : z_t = \ell \}$ given by $\Ndist\left( \bm{y}^i \mid \bm{1}\mu , \sigma^2 \bm{W} \right)$, where $\bm{W} = ( \kappa \bm{R}(\psi)^{ii} + \bm{I} )$ and $\bm{I}$ is the conforming identity matrix. Keep in mind that $\bm{W}$ is dependent on $\psi$. Now let $\hat{\mu} = ( \bm{1}' \bm{W}^{-1} \bm{1} )^{-1} \bm{1}' \bm{W}^{-1} \bm{y}^i = \sum_{j = 1}^{n_\ell} ( \bm{W}^{-1} \bm{y}^i )_j / w $ where $ w = \bm{1}' \bm{W}^{-1} \bm{1} $, and $s = ( \bm{y}^i - \bm{1}\hat{\mu} )' \bm{W}^{-1} ( \bm{y}^i - \bm{1}\hat{\mu} )$. The joint density for $\bm{y}^i$ can then be factored as
\begin{align}
	\label{eq:margy}
	p( \bm{y}^i \mid \cdots, -\bm{f} ) \propto \det(\bm{W})^{-1/2} \ ( \sigma^2 )^{-n_\ell / 2} \exp \left[ -\frac{ w(\hat{\mu} - \mu)^2 + s }{2 \sigma^2} \right] \, .
\end{align}
Now using the prior for $\mu$, we can further marginalize to obtain
\begin{align}
	\label{eq:margy2}
	p( \bm{y}^i &\mid \cdots, -\bm{f}, -\mu ) \propto \int p( \bm{y}^i \mid \cdots, -\bm{f} ) \ \Ndist(\mu \mid m_0, v_0) \diff \mu \nonumber \\
	&\propto \det(\bm{W})^{-1/2} \ ( \sigma^2 )^{-n_\ell / 2} \exp \left[ -\frac{ s }{2 \sigma^2} \right] \left( \frac{\sigma^2}{w} \right)^{1/2} c \, ,
\end{align}
where
\begin{align}
	\label{eq:margyC}
	c_0 &= \int \Ndist( \hat{\mu} \mid \mu, \sigma^2 / w ) \ \Ndist(\mu \mid m_0, v_0) \diff \mu \nonumber \\
	&\propto (\sigma^2/w + v_0)^{-1/2} \ \exp \left[ -\frac{ (\hat{\mu} - m_0)^2 }{ 2(\sigma^2/w + v_0) }  \right] \ \int \Ndist(\mu \mid m_1, v_1) \diff \mu \nonumber \\
	&= (\sigma^2/w + v_0)^{-1/2} \ \exp \left[ -\frac{ (\hat{\mu} - m_0)^2 }{ 2(\sigma^2/w + v_0) }  \right] = c \, ,
\end{align}
with $v_1 = \left( v_0^{-1} + w / \sigma^2 \right)^{-1}$ and $m_1 = v_1 ( m_0 / v_0 + w \hat{\mu} / \sigma^2 ) $.

A full Gibbs scan for $(\mu, \sigma^2, \bm{f}, \kappa, \psi)_\ell$ then proceeds as follows:
\begin{enumerate}
	\item Perform a random-walk Metropolis update of $(\kappa, \psi)$ with their joint collapsed conditional density proportional to $p( \bm{y}^i \mid \cdots, -\bm{f}, -\mu ) \ p(\kappa \mid \nu_\kappa, \kappa_0) \ p(\psi \mid \nu_\psi, \psi_0)$ where the first density is given in (\ref{eq:margy2}) and the remaining two are the inverse-gamma densities in (\ref{eq:mixcompFCall}). Gaussian proposals are drawn on the logarithmic scale, requiring a Jacobian adjustment by multiplying the collapsed conditional density by $\kappa \psi$ when computing the acceptance probability.
	\item Draw $\mu$ from its collapsed conditional distribution $\Ndist(m_1, v_1)$.
	\item Draw $\sigma^2$ from its collapsed conditional with density proportional to \\ $p( \bm{y}^i \mid \cdots, -\bm{f} ) \, p(\sigma^2 \mid \nu_\sigma, s_0)$, where the first density is given in (\ref{eq:margy}) and the second is the inverse-gamma density in (\ref{eq:mixcompFCall}). The result is another inverse-gamma density with shape $(\nu_\sigma + n_\ell)/2$ and scale $ (\nu_\sigma s_0 + w(\hat{\mu} - \mu)^2 + s)/2$.
	\item Introduce $\bm{f}^i$ with $\bm{f}^o$ still marginalized and draw from $p(\bm{f}^i \mid \cdots, -\bm{f}^o) \propto \Ndist(\bm{y}^i - \bm{1}\mu \mid \bm{f}^i, \sigma^2 \bm{I} ) \ \Ndist(\bm{f}^i \mid \bm{0}, \kappa \sigma^2 \bm{R}^{ii} ) $, a standard conditionally conjugate multivariate Gaussian update with covariance matrix $\bm{\Sigma} = \sigma^2 \left( \kappa^{-1} (\bm{R}^{ii})^{-1} + \bm{I} \right)^{-1}$ and mean vector $\bm{\Sigma}\left( \bm{y}^i - \bm{1}\mu \right) / \sigma^2 $. Following \citet[p.\ 46]{rasmussen2006gp}, the positive definite matrix $\bm{\Sigma}$ is computed, using the matrix inversion lemma, as $\sigma^2 \bm{K}(\bm{I} - \tilde{\bm{K}})$ where $\bm{K} = \kappa \bm{R}^{ii}$ and $\tilde{\bm{K}}$ is the solution to $(\bm{K} + \bm{I}) \tilde{\bm{K}} = \bm{K}$.
	\item Finally, draw $\bm{f}^o$ from its full conditional distribution. Let $\bm{C} = \kappa \sigma^2 \bm{R} $. Then we have $p(\bm{f}^o \mid \cdots) = \Ndist \left[ \bm{f}^o \mid \bm{C}^{oi} (\bm{C}^{ii})^{-1} \bm{f}^i, \bm{C}^{oo} - \bm{C}^{oi} (\bm{C}^{ii})^{-1} \bm{C}^{io} \right]  $.
\end{enumerate}


\section{SBM-multinomial update} 
\label{sec:appendix_SBM}

We describe the conjugate update for an SBM-distributed probability vector with a multinomial sampling model from \citet{heiner2019spv}, applied to Step 2 of the Gibbs sampler in Section \ref{sec:MCMC}.

Consider a length-$N$ sequence of independent random variables $\{z_t\} \in \{0, 1, \ldots, L\}^N$ with common distribution $\bm{\lambda} = (\lambda_0, \lambda_1, \ldots, \lambda_L)$. Given $\bm{\lambda}$, the probability of the sequence is $\prod_{t} \lambda_{z_t} = \lambda_0^{n_0} \cdots \lambda_L^{n_L}$ where the sufficient statistics in $\bm{n} = (n_0, \ldots, n_L)$ count the occurrences of each category.

Suppose $\bm{\lambda}$ follows the SBM distribution with parameters $\pi_{1}$, $\pi_{3}$, $\eta$, $\{\gamma_j\}$, and $\{\delta_j\}$. Let 
\begin{align*}
	g_j(a_j, b_j, \bm{n}) \equiv \frac{\Gamma(a_j + b_j)}{\Gamma(a_j^* + b_j^*)} \frac{\Gamma(a_j^*)}{\Gamma(a_j)} \frac{\Gamma(b_j^*)}{\Gamma(b_j)} \, ,
\end{align*}
with $a_j^* \equiv a_j + n_j$, and $b_j^* \equiv b_j + \sum_{h=j+1}^L n_h \, $, for $j = 0, 1, \ldots, L$. Then the marginal distribution of $\{ z_t \}$ has probability mass function
\begin{align}
	\label{eq:SBMmarg}
	p(\{z_t\}) = \prod_{j=0}^{L-1} [ &\pi_{1} \, g_j(1, \eta, \bm{n}) + \pi_{2} \, g_j(\gamma_j, \delta_j, \bm{n}) + \pi_{3} \, g_j(\eta, 1, \bm{n}) ] \, 
\end{align}
over its support, where $\pi_2 = 1 - \pi_1 - \pi_3$.

Now considering the update for $\bm{\lambda}$ in the GPMTD model, we have
$p(\bm{\lambda} \mid \cdots) \propto p(\bm{\lambda})\, \prod_t p(z_t \mid \bm{\lambda}) = \SBM( \bm{\lambda}; \pi_1, \pi_3, \eta, \bm{\gamma}, \bm{\delta}) \, \prod_t \lambda_{z_t} $, a conjugate SBM-multinomial update using the counts of $z_t$ in each of $\{0,1, \ldots, L\}$. A draw from the full conditional distribution begins by drawing the latent stick-breaking weights $\theta_\ell$, for $\ell=0,\ldots,L-1$, each from a mixture of three beta distributions. The mixture weights for $\theta_\ell$ are the three summands in the corresponding product terms of (\ref{eq:SBMmarg}), where $n_\ell$ is the cardinality of $\{t:z_t = \ell \}$. The three beta distributions have the corresponding $a_\ell^*$ and $b_\ell^*$ shape parameters taken from the SBM prior parameters and counts. The draw for $\bm{\lambda}$ is then constructed from the sampled $\{ \theta_\ell \}$ using the stick-breaking construction (\ref{eq:stickbreaking}).

\end{document}